\documentclass[aps,prl,twocolumn,showpacs,amssymb]{revtex4}

\usepackage{graphicx}
\usepackage{dcolumn}
\usepackage{bm}
\usepackage{color}

\def\be{\begin{equation}}
\def\en{\end{equation}}
\def\bea{\begin{eqnarray}}
\def\ena{\end{eqnarray}}
\def\bec{\begin{equation}\begin{array}{rcl}}
\def\p{\partial}

\def\gs{\gtrsim}
\def\ls{\lesssim}
\def\ve{\varepsilon}

\newcommand{\bi}[1]{\mbox{\boldmath$#1$}}

\newcommand{\ppp}[3]{{\bigg(}\frac{\partial {#1}}{\partial {#2}}{\bigg )}_{#3}}

\begin{document}
\title{Electric double layer  composed of 
 an antagonistic salt in an aqueous  mixture:\\ Local charge separation 
and surface phase transition  }
\author{Shunsuke Yabunaka$^a$ and Akira Onuki$^{b}$ }
\address{
$^a$ Fukui Institute for Fundamental Chemistry, Kyoto University, Kyoto 606-8103, Japan \\
$^b$ Department of Physics, Kyoto University, Kyoto 606-8502,
Japan\\
}


\date{\today}

\begin{abstract}
We examine  an electric double layer containing  an antagonistic salt 
 in an aqueous mixture, where   the cations are  
small  and hydrophilic but  the anions 
are large and  hydrophobic. In this situation, a  strong 
 coupling  arises between the  charge density and the solvent composition. 
As a result, the anions  are 
  trapped  in an oil-rich   adsorption layer on a hydrophobic  wall.
We then vary the surface  charge  density  $\sigma$ on the wall. 
For $\sigma>0$ the anions remain accumulated, 
but  for $\sigma<0$   the cations are  attracted 
to the wall with increasing $|\sigma|$. 
 Furthermore, the 
electric potential drop $\Psi(\sigma)$ is     
 nonmonotonic when the solvent interaction parameter $\chi(T)$ 
exceeds a critical value $\chi_c$  determined by  the composition  
and the ion density in the bulk. 
This  leads  to  a first order phase transition 
between two kinds of electric double layers 
with different $\sigma$ and  common $\Psi$. 
In  equilibrium  such two layer regions 
can coexist.  The steric effect due to finite ion sizes 
is crucial in these phenomena. 
\end{abstract}


\pacs{ 61.20.Qg, 68.05.Cf, 82.60.Lf, 82.65.Dp }
\maketitle


The electric double layer at a solid-liquid  interface is one of the most 
important entities in physical chemistry  
\cite{Is,Butt,Ben1,Bazant}. Its various aspects have long been studied 
mostly for one-component solvents 
with      the  mean-field Poisson-Boltzmann approach. 
However,  in a mixture solvent,  the ions   interact with 
the two solvent components   differently, leading 
 to a coupling between the charge density and the solvent 
composition\cite{Ben,Onuki,Bu,Tsori,Roij,Leibler,Oka}. 
This coupling is amplified in an aqueous mixture 
when the salt is composed of  hydrophilic  and  
  hydrophobic ions (antagonistic salt)
\cite{Onukireview,Nara,Ciach,Sada,SadaPRL,Leys}. 
In liquid water,  small hydrophilic    ions such as Na$^+$ 
 are surrounded by several  water  molecules   
  due to the ion-dipole  interaction\cite{Is}. 
A notable  example of hydrophobic ions  
is  tetraphenylborate BPh$_{4}^-$, which    
consists of  four phenyl rings bonded to an  ionized 
boron\cite{tetra}.  Because of its large size, it largely deforms  
the surrounding hydrogen bonding \cite{Garde,Chandler}. 
On the other hand, the ion solvation in nonaqueous solvent  
remains not well understood.

When hydrophilic and hydrophobic ions 
are added  in an aqueous  mixture,  local charge separation 
occurs   in the presence of compositional  heterogeneity. 
Indeed, in a x-ray  reflectivity experiment, Luo {\it et al}.\cite{Luo} 
 observed such ion distributions around a water-nitrobenzene 
interface. The resultant  double layer reduces the surface 
tension \cite{Onuki,Bu,Nara}, as has been observed \cite{Bonn}. 
Adding a small amount of   NaBPh$_{4}$ 
in  D$_2$O-trimethylpyridine, Sadakane {\it et al.} 
found a mesophase near its  criticality  \cite{Sada,Leys} 
and   multi-lamellar (onion) 
structures    far from it \cite{SadaPRL}. 

The interactions of  large 
hydrophobic ions with various soft matters  
are strong and  sometimes dramatic \cite{Leon,Faraudo2}. 
As an example,  Calero {\it et al.}\cite{Faraudo1} numerically  studied  
accumulation  of   BPh$_{4}^-$ 
 near  a  wall  in pure water solvent  to   explain 
a charge inversion effect  of  colloidal particles.  
In the presence of  a positive surface charge,  
they found that the  BPh$_{4}^-$ density was peaked at 
a short distance of $2.5~{\rm \AA}$ for a hydrophobic wall, 
while it   was  broadly peaked  at $3$ nm  for a hydrophilic wall. 

In  an aqueous mixture,   
hydrophilic (hydrophobic) ions are  
selectively adsorbed into a water-rich (oil-rich) adsorption layer 
 \cite{Onukireview}. 
In this Letter, we  further 
examine the distributions of    
hydrophobic anions (BPh$_{4}^-$) and  hydrophilic cations (Na$^+$) 
 next to     a hydrophobic  wall varying the surface charge density 
 $\sigma$.  For   $\sigma\ge 0$, 
  the  anions  remain  accumulated in the adsorption 
layer.  However, for   $\sigma< 0$, 
 the  cations   are eventually attracted to the wall 
 with increasing $|\sigma|$, where 
 the  composition profile  also  changes. 
 We shall see that this changeover   
 takes  place  as a first-order phase transition 
in some conditions of the parameters 
in our model.  We treat large hydrophobic anions, 
so  we should also account for 
 the steric effect due to finite ion sizes.  
This  effect   has been studied  in several  
papers in different situations  
\cite{Stern,Bike,Bie,Bazant,Ig,Andel,Maggs,Sha}.

As in Fig.1, we consider an electric double layer 
on a metal surface  at $z=0$.  The $z$ axis is perpendicular to 
the surface.  
The solvent consists of a waterlike component (called water)  
and a less polar component (called oil) 
with densities  $n_{\rm w}$ and $n_{\rm o}$, respectively.
For simplicity, they have the same molecular volume 
$v_0$, so  their volume fractions are $\phi=v_0 n_{\rm w}$ 
and $\phi'= v_0 n_{\rm o}$. 
The cations and anions are   monovalent  with densities 
 $n_1$ and $n_2$, respectively.
 Far from the wall,  we have 
$ n_1\to n_0$, $n_2\to n_0$,  and $\phi\to \phi_\infty$. 
We set  $n_0= 4\times 10^{-3}v_0^{-1}$ and vary $\phi_\infty$. 
Space is measured 
in units of $a\equiv v_0^{1/3}(\sim 3~{\rm \AA} )$  
and the Boltzmann constant is unity.

Introducing effective 
cation and anion  volumes  $v_1$ and  $v_2$, we assume   
the total volume fraction is unity:  
\be 
\phi+ \phi'+  v_1 n_1 +v_2 n_2=1 ,
\en 
which  holds for very small compressibility. 
 For  polymer mixtures the  
space-filling condition in the same form 
has been assumed \cite{Flory}. 
In our case we take $v_0 $ as the inverse density 
in a one-component liquid  of the first species  
 at given  $T$ and $p$ (for example, 
water at   $300$ K and $1$ atm). Around this reference 
 liquid we may    define 
$v_i$  in the dilute  limit of ions ($n_1\to 0$ and $n_2\to 0$ 
at  $n_{\rm o}=0$)  as    
\be 
\frac{v_i }{v_0} = - \ppp{n_i}{ n_{\rm w}}{ Tp n_j}^{-1} \quad (i=1,2, j\neq i).\en  
This ratio   is    also written as $(\p p/\p n_i)/( \p p/\p n_{\rm w})$, where 
   $p$ depends   on the densities and  $T$. 
We  assume  that Eq.(1) is a good approximation 
even for not small   $v_i n_i$ (up to $0.2$ in our analysis)  
 at fixed $T$ and $p$ \cite{comment}.  
See Supplemental Information (SI)  \cite{Supp}.   At present, 
 we have  no experimental data 
of  $v_i$ from Eq.(2), so 
we  set   $v_1/v_0=0.5$ for small cations 
\cite{Is,Hung,Marcus} and $v_2/v_0=5$  for  large anions \cite{tetra}.

\begin{figure}[tbp]
\begin{center}
\includegraphics[width=220pt]{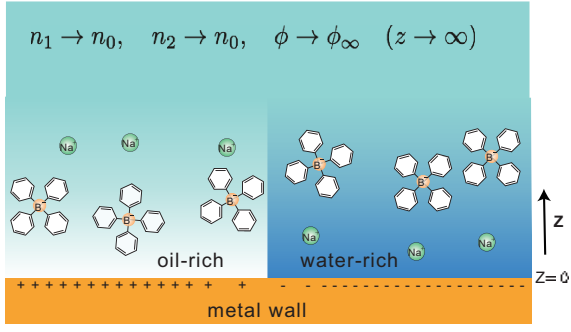}
\caption{(Color online) Electric double layer containing  
small hydrophilic cations (Na$^+$) and large 
hydrophobic anions (BPh$_4^+$) in an aqueous  mixture   
on a metal wall. There can be two kinds of  ion distributions  
with a common potential drop $\Psi$, which coexist 
in  certain conditions (see  Figs.4-6). 
Gradation of solvent region represents  water concentration.  
}
\end{center}
\end{figure}

\begin{figure}[tbp]
\begin{center}
\includegraphics[width=240pt]{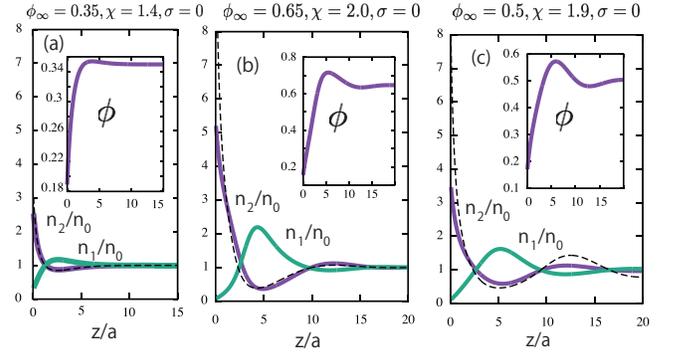}
\caption{(Color online) Hydrophilic cations $n_1(z)$ and 
 hydrophobic anions $n_2(z)$ 
  next to  hydrophobic wall  
for  $\sigma=0$ with $n_0=4\times 10^{-3}
v_0^{-1}$. Bulk water composition 
$\phi_\infty$ and interaction parameter 
$\chi$ are  (a) $0.35$ and $1.4$, 
(b) $0.65$ and $2.0$,  and (c) $0.5$ and $1.9$.
Water profile  $\phi(z)$ is also shown (insets).   
Anions are  richer in the oil-rich adsorption layer.  
Steric effect due to finite sizes of ions 
are accounted for (bold lines). Broken lines represent  anion profiles  
without steric effect ($v_1=v_2=0$).   
}
\end{center}
\end{figure}

The bulk  free energy density is  given by 
 \cite{Oka,Onuki,Tsori,Ben,Bu,Onukireview,Nara,Ciach,Ben1,Leibler,Roij}  
\bea 
f &=& \frac{T}{v_0}( \phi\ln \phi + \phi'\ln\phi'+\chi\phi\phi') + 
\frac{1}{2}C|\nabla\phi|^2  \nonumber\\ 
&&\hspace{-9mm}+ \sum_{i=1,2}n_i[T \ln (n_iv_i) -T +  
{\mu_{\rm sol}^i(\phi)} ]  +   
\frac{\ve(\phi)}{2}|{\bi E}|^2,   
\ena   
where   $\chi$ is the interaction parameter  depending on $T$ and 
we set  $C=  T/a$ \cite{Onukibook,Safran}. 
The $\mu_{\rm sol}^i$ is the solvation chemical potential, 
which is negative (positive) 
for hydrophilic (hydrophobic) ions. 
Its difference $\Delta\mu_{\rm sol}^i$ between coexisting two phases is  
 the Gibbs transfer free energy (per ion), whose size  
is large ($\gg T)$ in aqueous mixtures  in strong segregation 
\cite{Hung} but  is of order $T$   
for water-alcohol  in weak segregation \cite{Marcus}. Here, we assume the linear form,  
\be 
\mu_{\rm sol}^i(\phi)= -Tg_i \phi, 
\en 
with  $g_1= -g_2=10$.  
Then, $\Delta\mu_{\rm sol}^i \sim \pm 10T$ for strong segregation \cite{Hung}. 
The last term in $f$ is the electrostatic part, where 
 $\ve$ is the  dielectric constant and 
${\bi E}= -\nabla\psi$ is  the electric 
field. We assume the 
linear form  $ \ve(\phi)= \ve_0+ \ve_1\phi$ \cite{Debye} 
with  $\ve_0=\ve_1= e^2/12\pi a T$.  The Bjerrum length 
is then $3a/(1+\phi)$. 
Most previous papers  treated 
the simple case  $v_1=v_2=v_0$ \cite{Ig,Bazant,Bike,Andel}, 
but some attempts  were also made  
for  the asymmetric case $v_1\neq v_2$ 
    \cite{Maggs,Bie}.

The surface free energy density at $z=0$ 
is of the simple form  $f_s= h_1\phi$, 
 where $h_1$ is the surface field arising from 
the solvent-wall interactions 
\cite{Roth}. Minimizing 
the total free energy  $F= \int_{z>0} d{\bi r} f + 
\int_{z=0} dx dy  f_s$ \cite{Roth}, we find   
the boundary condition $\p\phi/\p z =h_1/C$ at $z=0$.
Supposing  a hydrophobic wall, we  set $h_1=  0.2T/a^2$ 
to obtain $\phi(z)=\phi(0)+ 0.2 z/a+\cdots $ for small $z$.

The electric potential $\psi$ obeys the  Poisson equation 
$\nabla\cdot\ve\nabla\psi= e ( n_2-n_1)$, where 
  $\psi \to  0$ as $z\to \infty$. Then,  $\Psi\equiv  \psi(0)$ 
 is the  potential drop  across the layer, 
which is independent of $ (x,y)$ on a metal surface. 
In this Letter, we  control the  surface charge 
$Q= \int dx dy~  \sigma$, where $\sigma(x,y)$ 
is the  charge density    related to $\psi$ by 
\be 
\sigma= -  \ve \p \psi/\p z \quad (z=0). 
\en 
We calculated  all the profiles assuming 
homogeneity of  the chemical potentials 
$\mu_\phi= \delta F/\delta \phi$ and $\mu_i =  \delta F/\delta n_i$ 
together with the Poisson equation for $z>0$. 
Here, $\mu_\phi$ and $\mu_i $ are 
 determined by  $\phi_\infty$ and $n_0$ 
(see their explicit expressions in SI \cite{Supp}).

We are not very  close to  the solvent criticality 
($\chi= 2$ and $\phi_\infty= 0.5$) in the bulk. 
In its vicinity,  a mesophase appears  in the bulk 
with  addition of 
an antagonistic salt \cite{Ciach,Sada,Onuki,Onukireview,Nara}. 
We are  also away from the solvent coexistence curve 
limiting ourselves to  the case   $\chi \le 2$, 
so we do not discuss   the wetting with ions 
\cite{Roij,Bu,Oka}.  
In this situation, we first  seek  one-dimensional (1D)  profiles 
fixing $\sigma$, where all the quantities depend only on $z$. 
 For constant    $\mu_\phi$ and $\mu_i$, 
 we consider   the grand potential density, 
\be 
\omega=\int_0^\infty dz[ f - 
\mu_\phi\phi-\sum_i \mu_i n_i + p_\infty ] + h_1 \phi(0).
\en 
where $p_\infty= \mu_\phi\phi_\infty +\sum_i \mu_i n_0 -f(\infty)$. 
We then find  
\be 
d\omega/d\sigma= \Psi,   
\en 
at fixed   $n_0$ and $\phi_\infty$ (see its derivation in SI \cite{Supp}). 
 Thus $\Psi$ is the  field variable conjugate to 
  $\sigma$. We require  $d\Psi/d\sigma>0$  for 
the thermodynamic stability.

For $\sigma=0$,   local charge separation occurs   
due to the presence of an  oil-rich adsorption layer on a hydrophobic wall. 
In Fig.2, the anions   accumulate   
    for $z< \ell_1 \sim 3a$, while 
 the cations are richer  in the next layer $\ell_1 <z<\ell_2 \sim 7a$. 
In (a), it is  relatively  mild   with  $\phi_\infty=0.35$ and $\chi=1.4$, 
 where the solvent  is   oil-rich at any $z$. 
However, it  is  more amplified in (b) and (c).  
Indeed, the deviation $\phi_\infty-\phi(z)$ is enlarged 
 with   $\phi_\infty=0.65$  and $\chi=2$ in (b), 
while  the criticality is  closer  with 
$\phi_\infty=0.5$ and $\chi=1.9$ in (c). 
The normalized potential drop $e\Psi/T$ 
is  (a) $-0.40$,   (b)  $-2.31$, and  (c) $-1.16$. 
Furthermore, in  (b) and (c), the deviations of 
 $\phi$, $n_i$, and $\psi$  are  strongly coupled  
even in the bulk, leading to  
 oscillatory decays (as  a precursor 
 of the mesophase)\cite{Oka,Ciach}.
In addition, $v_2n_2(0) \sim  0.1 $ in (b).  
Thus, to check relevance of the steric effect, 
 we also calculated  $n_2(z)$  for 
 $v_1=v_2=0$ \cite{Onukireview}. The resultant  $n_2(0)$ 
at $z=0$   is twice larger 
than that  with the steric effect 
 in (b) and (c), but is larger only by   $20\%$  in (a).   
Notice that  neutral colloidal particles   in the same 
situation  behave as negatively charged 
particles \cite{Faraudo1}.  

\begin{figure}[tbp]
\begin{center}
\includegraphics[width=240pt]{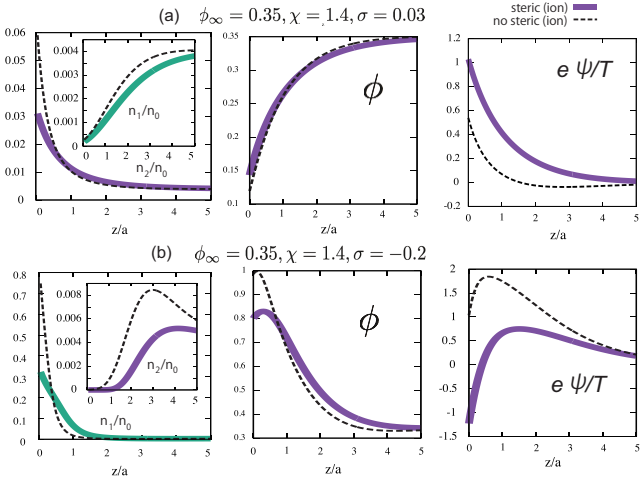}
\caption{(Color online)   1D profiles 
of ions  (left),  $\phi(z)$ (middle), and $e\psi(z)/T$ (right) 
for  $\sigma=0.03$ (top) and $-0.2$  (bottom) near a hydrophobic  wall, 
where $\phi_\infty=0.35$ and $\chi=1.4$. These states are stable 
on the curve in  Fig.4(a). Broken lines are obtained without 
 steric effect  ($v_1=v_2=0$).
 }
\end{center}
\end{figure}

\begin{figure}[t]
\begin{center}
\includegraphics[width=240pt]{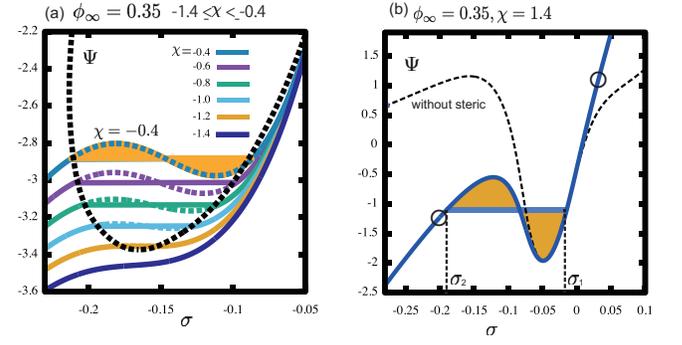}
\caption{(Color online)  Potential drop 
$\Psi$ vs  surface charge density $\sigma$ 
(in units of $T/e$ and $e/ a^2$, respectively)   
 from 1D solutions for $\phi_\infty =0.35$, where  
 (a) $\chi= -0.4, -0.6, -0.8, -1, -1.2$, and $-1.4$ and  (b) $\chi=1.4$. 
From  Eq.(9), first order phase 
transition occurs between  two layer states at $\sigma=\sigma_1$ 
and $\sigma_2$. Two colored  regions in yellow 
have the same area in (a) and (b).  
Two points ($\circ$) in (b) on the curve represent two states 
at $\sigma=-0.2$ and 0.03  in Fig.3. 
Dotted line in (b) represents 
 $\Psi$ without steric effect ($v_1=v_2=0$). 
}
\end{center}
\end{figure}

In Fig.3, we give  profiles of $n_i$, $\phi$, and $\psi$ 
for (a) $\sigma=0.03$ and (b) $\sigma= -0.2$, where  $\phi_\infty=0.35$ and 
$\chi=1.4$ (see Fig.4(b) for the corresponding 
states).  In (a), the anion accumulation is  stronger  than in Fig.2(a) 
(where $n_2(0)$ is 3 times larger) and the cations are expelled 
from the wall.  In (b), the  surface charge  density $-0.2$ 
is    largely negative, which is needed  to induce 
 cation accumulation at the hydrophobic wall. 
 In (b), we then 
find  $v_1 n_1(0) \sim   0.15$, where 
 $\phi(z)$  exceeds $ \phi_\infty$ 
at any $z$. Here,   $e\Psi /T$ is equal to (a) 
$1.0$  and (b) $-1.0$.

In Fig.4(a),  we show  $\Psi$ vs   $\sigma $ 
for several $\chi$ at $\phi_\infty=0.35$. Here, 
$\Psi$ has  local maximum and  minimum 
as a function of $\sigma$ for  $\chi> \chi_c 
=  -1.243$. Generally,   $\chi_c$ 
 depends  on  $\phi_\infty$ and $n_0$.  
This indicates  coexistence of two surface layers  
at $\sigma=\sigma_1$ and $\sigma_2$ 
with a common $\Psi(\sigma_1)=\Psi(\sigma_2)$  
for  $\chi >\chi_c$. Let  the areas of these   layers 
be  $S_1$ and $S_2$, where $S= S_1+S_2$ is 
the total wall area. At  fixed charge  $Q= S_1\sigma_1+ 
S_2\sigma_2$,  we   minimize the total grand potential, 
\be 
\Omega= S_1\omega(\sigma_1)+ S_2\omega(\sigma_2) - 
\lambda (S_1\sigma_1+S_2\sigma_2-Q),
\en 
with respect to $\sigma_1$, $\sigma_2$, and $S_1$. The   
   $\lambda$ is  the Lagrange multiplier. 
With the aid of Eq.(7) we 
 find $\lambda= \Psi(\sigma_1)=\Psi(\sigma_2) $ and 
$\omega(\sigma_1)- \lambda\sigma_1= \omega(\sigma_2)- \lambda\sigma_2$. 
These  yield 
\be 
\omega(\sigma_2)- \omega(\sigma_1)=
\int_{\sigma_1}^{\sigma_2} d \sigma 
\Psi(\sigma)=  \Psi(\sigma_1) (\sigma_2-\sigma_1). 
\en 
which is  a Maxwell  rule \cite{Max}. 
In (b), we then find   $\sigma_1=-0.19$  and 
$\sigma_2=-0.014$ for $\chi=1.4$. 
See SI  for results 
in the range  $\sigma_1<\sigma<\sigma_2$  \cite{Supp}.
 
\begin{figure}[tbp]
\begin{center}
\includegraphics[width=240pt]{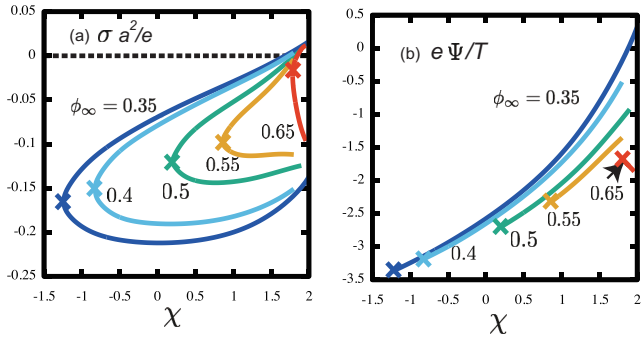}
\caption{(Color online)  Coexistence curves in 
(a) $\sigma$-$\chi$ plane and 
(b) $\Psi$-$\chi$ plane. 
for  $\phi_\infty=0.35, 0.4, 0.5, 0.55,$ and 0.65 at $n_0=4\times 10^{-3}
v_0^{-1}$.  For each $\phi_\infty$, two layers coexist inside 
the corresponding curve in (a), while $\Psi$ is a field 
variable  common in  coexisting two 
layers. Critical points are marked $(\times$).  
}
\end{center}
\end{figure}

In Fig.5, we display the coexistence curves in 
 the $\chi$-$\sigma$  and 
  $\chi$-$\Psi$ planes  for  
several $\phi_\infty$.  For each $\phi_\infty$, two layers coexist 
with $\sigma=\sigma_1$ and $\sigma_2$ inside 
the  corresponding curve in (a), while 
$\Psi$ is common in these layers in (b). 
Critical points are reached as $\sigma_2-\sigma_1\to 0$, 
 which form  a critical line on the coexistence  surface in the 
$\chi$-$\sigma$-$\phi_\infty$  
(or $\chi$-$\Psi$-$\phi_\infty$) space (at  fixed $n_0$). 
These phase behaviors are sensitive to $v_1$, $v_2$, and $h_1$, 
though the  transition itself exists even for $v_1=v_2=0$.

\begin{figure}[tbp]
\begin{center}
\includegraphics[width=240pt]{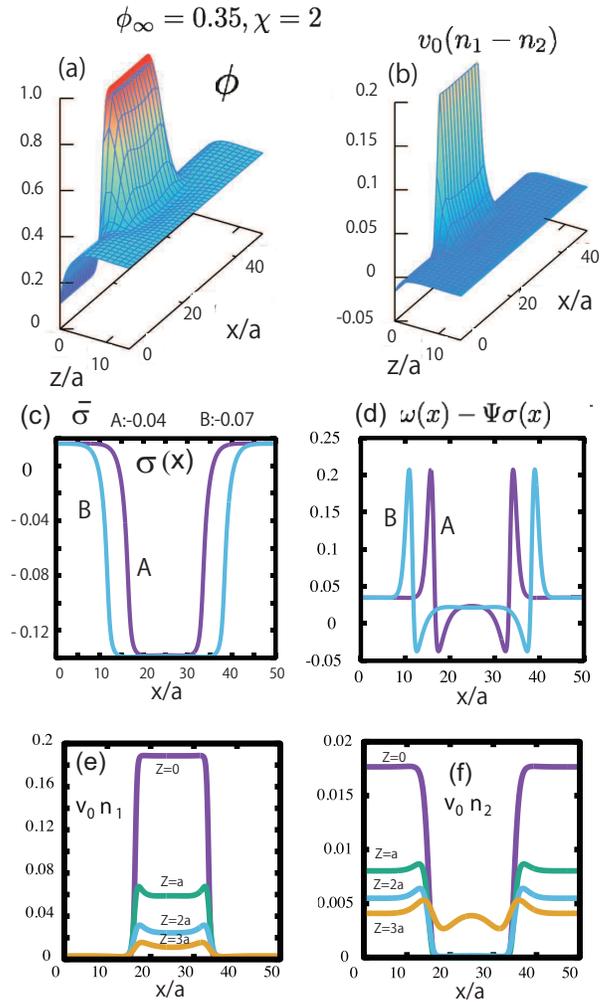}
\caption{(Color online)  Coexistence of two layers 
with $\sigma=0.02$ and $-0.14$  
for $\phi_\infty=0.35$ and $\chi=2$ (see Fig.5).
(a) $\phi (x,z)$ and (b) $ v_0(n_1(x,z)-n_2(x,z))$ on the $zx$ plane.   
(c) $\sigma(x) a^2/e $
 with  ${\bar{\sigma}}=Q/S $ being (A) $-0.04$ and (B)  $-0.07$ 
in units of $e/a^2$.   (d) $[\omega(x)- \Psi\sigma(x)]a^2/T$ 
for (A) and (B).  Cross-sectional profiles 
of  (e) $v_0n_1(x,a)$ and (f) $v_0 n_2(x,a)$ 
at $z/a=0,1,2$, and 3,  
 exhibiting small peaks at boundaries for $z\ge a$. 
}
\end{center}
\end{figure}

We calculated 2D  profiles from homogeneos 
$\mu_\phi$ and $\mu_i$  in the $zx$ plane with  
    $\chi=2$ and $\phi_\infty=0.35$. 
 For this $\phi_\infty$,  the phase transition behavior 
is not much changed in the range 
  $1.4\le \chi \le 2$ in Fig.5(a). In Fig.6, we 
show   (a) $\phi$ and (b) $v_0(n_1-n_2)$, where  
 a stripe region with $\sigma=\sigma_1=-0.14$ 
is embedded  between regions  with  $\sigma= \sigma_2=0.01$ at $\Psi=0.32$. 
Here,     the mean surface charge density 
${\bar\sigma}= Q/S$ is between $\sigma_1$ and  $\sigma_2$. 
In (c),    $\sigma(x)$   from Eq.(5) is  roughly equal to 
$\sigma_1$ or $\sigma_2$ except for the boundary regions. 
The fraction of the region with 
$\sigma=\sigma_1$ is nicely given by 
$(\sigma_2-{\bar\sigma})/(\sigma_2-\sigma_1)$. 
 In (d), we plot ${\hat\omega}(x)\equiv 
\omega(x)-\Psi \sigma(x)$, where $\omega$ is defined in Eq.(6). 
From Eq.(9) it  assumes 
a nearly common value $ {\hat\omega}_1$  in the two regions. 
The integral of  ${\hat\omega}(x)- {\hat\omega}_1$ 
across one of the boundary regions is the line tension $\tau$ \cite{Line}, 
which is of order  $0.1 T/a$ here. 
In (e) and (f), cross-sectional profiles of 
 $n_1$ and $n_2$ at constant $z$ are given, which 
 exhibit small peaks   
at the boundaries   slightly away from  the wall. 
This is  because of  the  Coulomb attraction between 
 the cations and the anions which are locally 
separated across the boundaries.  For  the same reason, 
 more  marked  peaks  appear in the densities of antagonistic ion pairs 
 near   water-oil interfaces 
 \cite{Luo,Onukireview,Onuki}.

We propose   experiments in the above situation. 
Let  $\bar\sigma$ be 
  decreased slightly below $\sigma_2$ 
on a hydrophobic metal wall. Then, the  oil-rich  layer with  hydrophobic 
anions  becomes metastable against formation of small  water-rich  
 regions with   hydrophilic cations. For  a finite line tension $\tau$, 
 their  shapes are  circular with  the critical radius \cite{Onukibook}  
\be 
r_c= \tau/[(d\Psi/d\sigma)(\sigma_2-{\sigma}_1)(\sigma_2-{\bar\sigma})],
\en  
where the derivative $d\Psi/d\sigma$ is taken at $\sigma=\sigma_2$. 
On the other hand,   a hydrophilic metal wall 
will be covered with  a water-rich layer  
for $\sigma\cong 0$, but   small 
oil-rich  regions  will be  nucleated 
with increasing $\sigma>0$.

In summary, we have found 
a first-order surface  transition with 
antagonistic ion pairs having   different  sizes. 
In future,  we should  examine wetting  near the solvent coexistence 
curve with  an antagonistic salt. We will  study behavior of 
colloidal particles (including Janus ones) 
in a mixture solvent with  
an antagonistic salt, where the ion distributions around them  
 can be  very complex.

\pagebreak
\widetext

{\bf Supplemental Information}\\
\vspace{2mm}
{\bf   Electric double layer  composed of 
 an antagonistic salt in an aqueous  mixture:\\ 
\hspace{5mm} Local charge separation 
and surface phase transition 
}\\
\vspace{2mm}
\hspace{3cm} 
 Shunsuke Yabunaka$^a$ and Akira Onuki$^{b}$\\
$^a$ Fukui Institute for Fundamental Chemistry, Kyoto University, Kyoto 
606-8103,  Japan\\
$^b$ Department of Physics, Kyoto University, Kyoto 606-8502,
Japan\\

\setcounter{equation}{0}
\renewcommand{\theequation}{S.\arabic{equation}}
\renewcommand{\thefigure}{S\arabic{figure}}

{\bf{ Space-filling condition and chemical potentials}}\\ 
We introduce the ion volumes $v_1$ and $v_2$ 
using their definittion in  Eq.(2) in our Letter. 
The total volume fraction $\phi_{\rm tot}$ is the sum of those of water, 
 oil, cations, and anions: 
\be 
\phi_{\rm tot}= \phi+\phi'+ v_1n_1+ v_2n_2.
\en 
where  $v_0$, $v_1$, and $v_2$ depend on $T$ and $p$ but 
not on the mole fractions of  the four components. 
We assume that  $\phi_{\rm tot}$ is  very  close to 1  
even for not very small $v_1n_1+ v_2n_2$. 
Its  deviation from 1 should  yield 
  an  increase in the Helmholtz free energy $\Delta F=\int_{z>0}d{\bi r}
\Delta f $ with   
\be 
\Delta f =  \gamma(\phi_{\rm tot}-1)^2/2v_0,
\en   
where $\gamma$ is a large coefficient ($\gg T$). If the fluid is homogeneous 
with volume $V$, the excess free energy is $\Delta F = 
\gamma (V-V_{\rm tot})^2/2Vv_0$ with $V_{\rm tot}=V\phi_{\rm tot}= 
 v_0(N_{\rm w}+ N_{\rm o})+ 
v_1 N_1+v_2N_2$, where $N_\alpha$ ($\alpha={\rm w},\rm{o},1$, and 2) 
are the total particle numbers. Its differentiation 
with respect to $V$   at fixed $N_\alpha$ gives the excess 
pressure, 
\be 
\Delta p= \gamma (\phi_{\rm tot}-1)/v_0.
\en 
We treat physical states with $|\Delta p|\ll T/v_0$. 
  If  $\gamma \gg T$, 
  the isothermal compressibility  (at fixed molar fractions) 
  is nearly equal to $v_0/\gamma$. It is worth noting that 
the compressibility of ambient  liquid water 
(300 K and 1 atm) is  $4.5\times 10^{-4}/$MPa $\sim 0.06 v_0/T$ 
for $a=v_0^{1/3}=  3~{\rm \AA}$.  

If we allow small deviations of  the space-filling condition (1), 
we  should replace  the total free energy $F=\int_{z>0}  d{\bi r} f+ 
\int_{z=0}  dxdy~ h_1\phi$  by $F+\Delta F$. 
With the aid of  Eqs.(3) and (4), the  chemical potentials are defined by 
${\hat\mu}_\alpha= 
\delta (F+\Delta F) /\delta n_\alpha$ ($\alpha={\rm w},\rm{o},1$, and 2), 
where $F+\Delta F$ 
is treated as a functional of $n_{\rm w}$, $n_{\rm{o}}$,  $n_1$, 
and $n_2$ at fixed $T$ and surface charge $Q=\int dxdy\sigma$. 
To calculate  these quantities 
we  consider  small variations 
 $\delta\phi$, $\delta n_i$, and $\delta\sigma$. Using   
 $\delta (\ve |{\bi E}|^2) 
= -|{\bi E}|^2\delta\ve  +2{\bi E} \cdot\delta(\ve {\bi E})$ 
and 
\be 
\int_{z>0} d{\bi r}{\bi E}\cdot{\delta (\ve{\bi E})}= 
\int_{z>0} d{\bi r}~\psi e(\delta n_1-\delta n_2)
+ \int_{z=0} dxdy~ \psi \delta (\ve E_z),
\en  
we obtain  the incremental change in $F+\Delta F$  as  
\be 
\delta (F+\Delta F)=\int_{z>0}  d{\bi r}\sum_{\alpha={\rm w},\rm{o},1,  2 } 
{\hat\mu}_\alpha  \delta n_\alpha+ 
\int_{z=0}dxdy \bigg[ (h_1-C \p \phi/\p z)\delta\phi + 
  \psi \delta (\ve E_z)\bigg], 
\en   
Then, since   $\delta D_z= \delta\sigma$ 
and   $C \p \phi/\p z=h_1$ at $z=0$,   the second term in Eq.(S.5) 
simply becomes $\Psi\delta Q$ on  a metal surface with $\Psi=\psi(0)$ 
for $\phi(\infty)=0$.  Some calculations give 
\bea
{\hat\mu}_{\rm w}&=&  T[\ln \phi +1  + \chi\phi'- v_0 \sum_i g_i n_i 
-  \frac{Cv_0}{T} \nabla^2\phi] 
- \frac{v_0 \ve_1}{2} |{\bi E}|^2  +\gamma (\phi_T-1), \\
{\hat\mu}_{\rm o}&=& T[ \ln \phi' +1  + \chi\phi]  +\gamma (\phi_T-1),\\
{\hat\mu_1}&=&  T[ \ln(n_1v_1)-g_1\phi]  + 
 {e\psi}  + \gamma (\phi_T-1)v_1/v_0, \\ 
{\hat\mu_2} &=&  T[\ln(n_2 v_2)-g_2\phi] - 
 {e\psi}  + \gamma (\phi_T-1)v_2/v_0. 
\ena 
For  equilibrium and  metastable profiles,  these chemical potentials 
are homogeneous constants. Using these  profiles,   we consider 
the grand potential defined by 
\be 
\hat{\Omega}=\int_{z>0} d{\bi r}\bigg[f+\Delta f + p_\infty  - 
\sum_\alpha {\hat\mu}_\alpha n_\alpha\bigg]+\int_{z=0} dxdy ~h_1\phi  ,
\en 
where   $p_\infty$  is a constant chosen to make 
the integrand in the first term vanish 
for large  $z$. Then, from $C \p \phi/\p z=h_1$ at $z=0$ and    Eq.(S.5), 
we obtain   
\be 
\delta \hat{\Omega}= \Psi\delta Q.
\en 
If $\hat\Omega$ is treated as a function of $Q$, 
we obtain $d{\hat\Omega}/dQ= \Psi$ for each 
$\phi_\infty$, $n_0$, and $T$. 
As  $\phi_{\rm tot} \to 1$ in 
the one-dimensional case, 
  $\hat\Omega$ in Eq.(S.10) tends 
to $S\omega$, where $\omega$ is defined by Eq.(6) 
and $S$ is the surface area of the metal wall.  
Then, we find  $d\omega/d\sigma=\Phi$ in Eq.(7). 

Below Eq.(5) of our Letter, we have 
introduced  $\mu_\phi=\delta F/\delta\phi$ and 
$\mu_i=\delta F/\delta n_i$ starting with Eq.(1) ($\phi_{\rm tot}=1$), 
where  $\phi'$ is eliminated and 
 $F$ is  a function of the three variables $\phi$, $n_1$, and $n_2$. 
For small $\phi_{\rm tot}- 1$ and large $\gamma/ T$, we  can express 
$\mu_\phi$ and $\mu_i$ as   
\be
\mu_\phi =( {\hat\mu}_{\rm w}-{\hat\mu}_{\rm o})/v_0,\quad~  
\mu_i={\hat\mu}_i- {\hat\mu}_{\rm o} v_i/v_0 \quad (i=1,2), 
\en 
where the terms proportional to 
$\gamma(\phi_{\rm tot}-1)$ are  eliminated. 


{\bf Changeover of layer profiles}\\
In   our Letter,  we have presented  numerical results 
for $\phi_\infty=0.35$ and $n_0=4\times 10^{-3}v_0^{-1}$ at $\chi=1.4$ 
  on a hydrophobic wall  in Figs.2-4. Here, adopting   these parameter values, 
we give 1D profiles of (a) $n_1(z)$, (b) $n_2(z)$, (c) $\phi(z)$, 
and (d) $\psi(z)$ in dimensionless units 
 in  Fig.S1. We set   $\sigma$ equal to (A) $0$,  
(B) $-0.04$,  (C) $-0.08$,  (D) $-0.12$,  and (E) $ -0.16$ in units of 
$e/a^2$, where $v_0=a^3$. 
 These quantities largely change with decreasing $\sigma$. 
In (a), the cations are  expelled from 
the wall for $\sigma=0$, 
but they abruptly  accumulate  near the wall  for $\sigma\ls -0.08$ 
because of their small size $v_1=0.5v_0$. 
In (b), the anions are accumulated near the wall with 
$v_2n_2(0) \cong 0.05$ for $\sigma=0$, but are 
expelled from the wall for  $\sigma\gs -0.08$. 
The  anions  accumulate more 
 weakly than the cations  because of their large size ratio 
$v_2/v_1=10$. In (c), the water volume fraction $\phi(z)$ is 
less than $\phi_\infty$ near the wall for $\sigma=0$ and -0.04, 
but is increased above $\phi_\infty$ for the lower  $\sigma$ values. 
In (d), the potential drop $\Psi=\psi(0)$ remains negative, 
but $\psi(z)$  gradually increases near the wall. For $\sigma\ls -0.08$, 
$\psi(z)$ exhibits a maximum at an intermediate $z_m\sim 1.5a$, 
so the electric field $E_z= -d\psi/dz$ is positive for $z<z_m$ and  
negative for $z>z_m$.

\begin{figure*}
\begin{center}
\includegraphics[width=450pt]{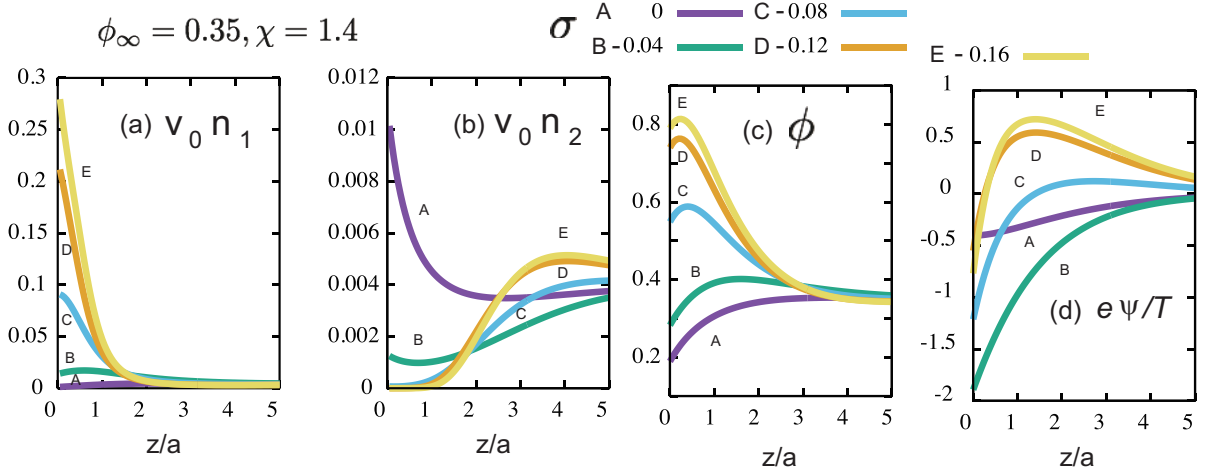}
\caption{(Color online) Changeover of profiles of 
(a) $v_0 n_1(z)$, (b) $v_0 n_2(z)$, (c) $\phi(z)$, 
and (d) $e\psi(z)/T$ for five values of $\sigma$, where $\chi=1.4$, 
 $\phi_\infty=0.35$,  and $n_0=4\times 10^{-3}v_0^{-1}$. }
\end{center}
\end{figure*}

\begin{figure*}
\begin{center}
\includegraphics[width=450pt]{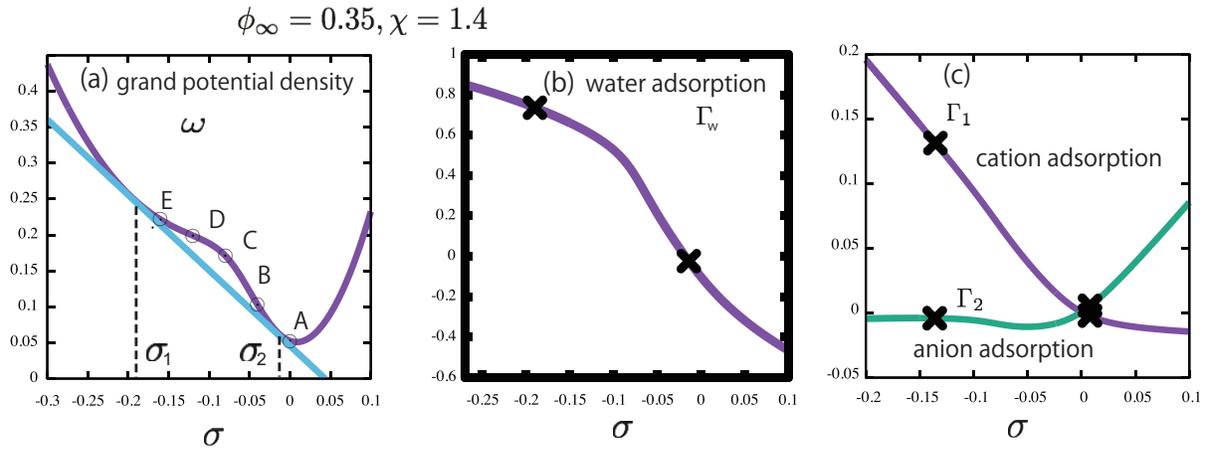}
\caption{(Color online) (a) $\omega a^2/T$, (b) $\Gamma_{\rm w} a^2$, 
and (c) $\Gamma_i a^2 $ 
($i=1, 2$) as functions of $\sigma$ (in units of $e/a^2$). 
First order phase transion occurs between two states 
at  $\sigma=\sigma_1=-0.19$ and  $\sigma=\sigma_2=-0.014$. In (a) 
points A, B, ...., and E correspond to those in Fig.S1. 
In (b) and (c) these points are marked by $\times$. 
The other parameter values are the same as those in Fig.S1. }
\end{center}
\end{figure*}

For the parameter values in Fig.S1, a 
 first order phase transition occurs between 
two surface charge densities given by  
 $\sigma_1=-0.19$  and $\sigma_2=-0.014$ from Fig.4(b).  
In Fig.S2(a), the grand potential density $\omega(\sigma)$ in Eq.(6)  
is plotted, where its tangential line at $\sigma=\sigma_1$ 
and that at $\sigma= \sigma_2$ coincide from Eq.(9) with a  common 
slope equal to the potential drop  $\Psi$. 
Thus, the state (A) is stable where $\sigma>\sigma_2$. 
However,  the states (B)-(E) are metastable or unstable 
because their  $\sigma$ values  are between $\sigma_1$ and $\sigma_2$. 
In (b) and (c), we plot the excess adsorbates   $\Gamma_{\rm w}$,  
 $\Gamma_{1}$, and $\Gamma_{2}$ for water molecules, 
cations, and anions, respectively. In our semi-infinite case 
they are defined by 
\be 
\Gamma_{\rm w}= v_0^{-1}\int_0^\infty dz [ \phi(z)- \phi_\infty],\quad\quad  
\Gamma_{i}= \int_0^\infty dz [ n_i(z)- n_0]\quad (i=1,2).
\en 
With decreasing $\sigma$,  $\Gamma_{\rm w}$ and $\Gamma_{1}$ increase, while 
  $\Gamma_{2}$ decreases to zero, which confirms the strong coupling between the composition and the ion densities.


\begin{thebibliography}{99}
\bibitem{comment} 
We may  use   the the Mansoori-Carnahan-Starling-Leland 
model for  high-density liquid mixtures \cite{Bazant,Bie,Maggs}  
to   justify Eq.(1), where  
$v_1$ and $v_2$ are  determined by 
the hard-core part of the free energy 
as functions of $T$ and $p$.
\bibitem{Safran} The coefficient $C$ in Eq.(3) depends on 
the microscopic interactions and can be determined 
from experimental data of the correlation length 
or the surface tension.
\bibitem{Max} For one-component fluids 
the chemical potential $\mu=\p f/\p n$ depends on the density $n$ 
at fixed $T$ where $f$ is the Helmholtz 
free energy density. When a gas with density $n_g$ and a liquid with 
density $n_\ell$ coexist, we have 
$\int_{n_g}^{n_\ell}dn \mu(n)= 
\mu(n_g)(n_\ell-n_g)$  analogously  to Eq.(9).
\end{thebibliography}
\end{document}